\documentclass[12pt, prd, showpacs]{revtex4}
\usepackage{amssymb}
\usepackage{amsmath}

\input{tcilatex}

\begin{document}

\title{Energy extraction from extremal charged black holes due to the BSW
effect}
\author{O. B. Zaslavskii}
\affiliation{Department of Physics and Technology, Kharkov V.N. Karazin National
University, 4 Svoboda Square, Kharkov, 61077, Ukraine}
\email{zaslav@ukr.net }

\begin{abstract}
Two particles can collide in the vicinity of a rotating black hole producing
the divergent energy in the centre of mass frame (the BSW effect).\ However,
it was shown recently that an observer at infinity can register quite modest
energies $E$ and masses $m$ which obey some upper bounds. In the present
work the counterpart of the original BSW effect is considered that may occur
even for radial motion of colliding particles near charged static black
holes. It is shown that in some scenarios there are no upper bound on $E$
and $m$ . Thus the high-energetic and superheavy products of the BSW effect
in this situation are, in principle, detectable at infinity.
\end{abstract}

\maketitle

The investigation of high-energetic particles collisions in the vicinity of
rotating black holes was initiated in \cite{ps} where the collisional
version of the Penrose process \cite{pen} was investigated. The new urge to
considering such processes came from an interesting observation made in Ref. 
\cite{ban}. It was found there that two particles which move towards the
horizon of the extremal black holes can produce an infinity energy in the
centre of mass frame $E_{c.m.}$. This effect (called the BSW one after the
names of its authors) provoked a large series of works and is under active
study currently. The most part of them was restricted to the investigation
of the vicinity of the horizon where collision occurs. Meanwhile, of special
interest is the question whether the consequences of this effect can be
detected (at least, in principle) in a laboratory. In other words, can an
observer at infinity register ultra-high energy and/or superheavy particles?
In \cite{flux1}, \cite{flux2} the phenomenological description of fluxes
from emergent particles due to collisions was suggested. The recent works 
\cite{p}, \cite{z}, \cite{j} discussed the collisional process when one of
two created particles escapes to infinity. It turned out that in spite of
divergent $E_{c.m.}$, the energy and mass which can be registered at
infinity as a result of the BSW effect are rather modest. In this sense, the
results of \cite{p}, \cite{z} and \cite{j} are somewhat disappointing in the
observational sense (although they do allow some indirect imprints of the
BSW effect on its products measured at infinity).

Meanwhile, there exists also the counterpart of the original BSW effect that
arises not due to rapid rotation but due to the charge of a black hole \cite%
{jl}. It reveals itself even for a pure radial motion. From another hand, it
is known that for the RN black hole the analogue of the Penrose process \cite%
{pen} also exists \cite{d} (details of a more general case when a black hole
is both rotating and charged can be found in the review \cite{nar}). In the
present work we consider the collisional versional of the Penrose process
for charged black holes and show that, contrary to the situation with
rotating black holes, the significant energy extraction is possible. Without
pretending to direct applications in realistic astrophysics, the example
with a charged nonrotating black hole can be viewed as a useful model that
shows how the BSW effect can have direct manifestations detectable at
infinity.

Let us consider the spherically symmetric metric%
\begin{equation}
ds^{2}=-N^{2}dt^{2}+\frac{dr^{2}}{N^{2}}+r^{2}(d\theta ^{2}+d\phi ^{2}\sin
^{2}\theta )
\end{equation}

describing the extremal Reissner-Nordstr\"{o}m metric (RN). Then, the lapse
function 
\begin{equation}
N=1-\frac{r_{+}}{r}  \label{N}
\end{equation}%
where $r_{+}$ is the horizon radius.

Let two particles 1 and 2 with masses $m_{1}$ and $m_{2}$ be injected from
infinity, collide and produce two new particles 3 and 4 with masses $m_{3}$
and $m_{4}$. We restrict ourselves by pure radial motion only
(generalization to nonzero angular momenta is straightforward). Then, the
conservation laws of the electric charge, energy and radial momentum give us%
\begin{equation}
q_{1}+q_{2}=q_{3}+q_{4}\text{,}
\end{equation}%
\begin{equation}
X_{1}+X_{2}=X_{3}+X_{4}\text{,}  \label{x}
\end{equation}%
\begin{equation}
\varepsilon _{1}Z_{1}+\varepsilon _{2}Z_{2}=\varepsilon
_{3}Z_{3}+\varepsilon _{4}Z_{4}\text{.}  \label{rad}
\end{equation}

Here, $X=E-q\varphi $ where $\varphi $ is the electric potential. $Z_{i}=%
\sqrt{X_{i}^{2}-m_{i}^{2}N^{2}}$. As usual, we assume that $X>0$ outside the
horizon (the forward in time condition) but $X_{H}=0$ is possible on the
horizon. Hereafter, subscript "H" means that the corresponding quantity is
calculated on the horizon. It is implied that $\varepsilon _{i}=-1$ if a
particle labeled by subscript "i" moves towards the horizon and $\varepsilon
_{i}=+1$ if it moves outwardly. For the extremal RN metric, $\varphi =\frac{%
r_{+}}{r}=1-N$ (for definiteness, the electric charge of a black hole $%
Q=r_{+}>0$).

The BSW\ effect occurs if particle (say) 1 is critical and particle 2 is
usual \cite{jl}. It means, by definition, that $\left( X_{_{1}}\right) _{H}=0
$, so $E_{1}=q_{1}$ and $\left( X_{2}\right) _{H}\neq 0$. Further
consideration goes very closely to \cite{j} but gives the results
qualitatively different from those for rotating uncharged black holes \cite%
{p}, \cite{z}, \cite{j}.

We must have $Z^{2}=X^{2}-m^{2}N^{2}\geq 0$, the zeros of $Z$ give us the
turning points. The condition $Z=0$ can be rewritten as $X=mN$, whence%
\begin{equation}
q\leq q_{0}=\frac{E-mN}{1-N}.  \label{q0}
\end{equation}%
On the horizon, $q_{0}=q_{H}\equiv E$. Hereafter, it is assumed that $q>0$
for a critical (or near-critical) particle, otherwise one cannot achieve the
condition of criticality $X_{H}=0$.

Near the horizon, (\ref{q0}) turn into%
\begin{equation}
q_{0}=E+N(E-m)+O(N^{2}).  \label{q0n}
\end{equation}

We will be interested in the situation when a particle (denoted as particle
3) escapes to infinity from the immediate vicinity of the horizon. This is
possible in 2 cases.

a) $E_{3}\geq m_{3}$, $q_{3}<q_{H}$, $\varepsilon _{3}=+1$. The act of
collision occurs in the allowed region just near the horizon, afterwards
particle 3 escapes to infinity.

b) $E_{3}\geq m_{3}$, $q_{H}<q_{3}<q_{0}$, $\varepsilon _{3}=+1$ or $%
\varepsilon _{3}=-1$. Collision occurs just outside the potential barrier
near the horizon, particle 3 is slightly noncritical. The condition $%
\varepsilon _{3}=-1$ means in this context that particle 3 is moving
inwardly, approaches the turning point and bounces back in the outward
direction. We consider all these types of scenarios in the vicinity of the
horizon where $N\ll 1$.

It is convenient to write $q=E(1+\delta ).$ Then, in case (a) $\delta <0.$In
case (b) $\delta \geq 0$ but it is bounded from the above. Indeed, forward
in time condition $X>0$ gives us%
\begin{equation}
\delta <\frac{N}{1-N}  \label{dn}
\end{equation}

The condition $q<q_{0}\,$\ entails%
\begin{equation}
\delta <N(1-\frac{m}{E})  \label{d-}
\end{equation}%
which is more tight than (\ref{d-}).

In the near-horizon region, the lapse function $N$ is a small quantity. For
what follows, we need also the expansions for the quantity $Z$. This can be
found separately for different kinds of particles.

1) Usual particle. For such \ a particle, $X_{H}\neq 0$, so we obtain

\begin{equation}
Z=X+O(N^{3})  \label{uz}
\end{equation}%
where%
\begin{equation}
X=X_{H}+qN\text{, }X_{H}=E-q\text{.}
\end{equation}

2) Critical particle. Now, $X_{H}=0$, $q=E$, so%
\begin{equation}
X=EN\text{,}  \label{cx}
\end{equation}%
\begin{equation}
Z=N\sqrt{E^{2}-m^{2}}\text{.}
\end{equation}

3) Near-critical particle

Let us consider a particle which is not exactly critical but, rather,
near-critical. For such a particle, $\delta \ll 1$. We can adjust the value $%
\delta $ to the small $N$ choosing it in the form of series%
\begin{equation}
\delta =C_{1}N+C_{2}N^{2}+...  \label{de}
\end{equation}

Then,%
\begin{equation}
X=EN(1-C_{1})+O(N^{2}),
\end{equation}%
\begin{equation}
Z=N\sqrt{E^{2}(1-C_{1})^{2}-m^{2}}+O(N^{2})  \label{znc}
\end{equation}

Now, we can apply the near-horizon expansion to different scenarios of
escaping In case (a), the condition $\delta <0$ and eq. (\ref{de}) give us%
\begin{equation}
C_{1}<0\text{.}  \label{cneg}
\end{equation}

In case (b), we must take into account the presence of the turning point
outside the horizon. Then, it follows from (\ref{d-}) that 
\begin{equation}
0\leq C_{1}\leq \left( C_{1}\right) _{m}=1-\frac{m}{E}.  \label{cc}
\end{equation}

The scenario in which a near-crticial particle has $\varepsilon _{3}=-1$
immediately after collision and thus moves inwardly, we will call IN
scenario for shortness. If after collision $\varepsilon _{3}=+1$ we will
call it "OUT" scenario. In turn, we will add superscript "-" if $\delta <0$
and "+" if $\delta \geq 0$. In other words, we enumerate possible types of
scenarios characterizing them by signs of two quantities - $\varepsilon $
and $\delta $. In general, there are 4 combinations: OUT$-$, OUT$+$, IN$+$.
and IN$^{\_}$. However, the scenario IN$-$ should be rejected since it
corresponds to a particle 3 falling down in a black hole whereas we want it
to escape to infinity..

In any scenario, particle 4 is usual and falls into a black hole ($%
\varepsilon _{4}=-1$). This is obtained in \cite{j} by analyzing eqs. (\ref%
{x}), (\ref{rad}) for the Kerr case but actually it is insensitive to the
form of the metric and relies on pure algebra, so it applies also to dirty
rotating black holes \cite{z} and to the RN one.

Using (\ref{rad}) with $\varepsilon _{1}=\varepsilon _{2}=\varepsilon _{3}=-1
$ we have$\ -Z_{1}-Z_{2}=-Z_{4}+\varepsilon _{3}Z_{3}$. Then, using (\ref{uz}%
), (\ref{cx}), (\ref{znc}) we obtain%
\begin{equation}
F\equiv A+E_{3}(C_{1}-1)=\varepsilon _{3}\sqrt{%
E_{3}^{2}(1-C_{1})^{2}-m_{3}^{2}}  \label{1st}
\end{equation}%
where%
\begin{equation}
A\equiv E_{1}-\sqrt{E_{1}^{2}-m_{1}^{2}}\text{,}
\end{equation}%
\begin{equation}
0\leq A\leq E_{1}\text{.}  \label{ae}
\end{equation}%
Taking the square of (\ref{1st}) we have%
\begin{equation}
C_{1}=1-\frac{m_{3}^{2}+A^{2}}{2E_{3}A}\text{,}  \label{ca}
\end{equation}%
\begin{equation}
F=\frac{A^{2}-m_{3}^{2}}{2A}\text{.}  \label{f}
\end{equation}%
It is seen from (\ref{ca}) that eq. (\ref{cc}) is satisfied automatically.
According to (\ref{1st}), $signF=\varepsilon _{3}$. Now we will discuss
different scenarios separately. 

OUT$-$  $\varepsilon _{3}=+1$, $C_{1}<0$

It follows from (\ref{ca}) and (\ref{f}) that%
\begin{equation}
m_{3}\leq E_{3}\leq \mu =\frac{m_{3}^{2}+A^{2}}{2A}.
\end{equation}%
\begin{equation}
m_{3}\leq A\text{,}  \label{ma}
\end{equation}%
\begin{equation}
\mu \leq A\leq E_{1}\text{.}
\end{equation}%
There is no energy extraction in this case. In particular, if $E_{1}=E_{2}=m$%
, the efficiency of extraction $\eta =\frac{E_{3}}{E_{1}+E_{2}}\leq \frac{1}{%
2}.\,$\ Moreover, the quantity $\mu $ is a monotonically decreasing function
of $E_{1}$. Scenario OUT$-$ was considered in the 1st version of preprint 
\cite{z} but two other scenarios which are the most interesting were
overlooked there. It is their consideration which we now turn to.

OUT+ $\ \varepsilon _{3}=+1$, $C_{1}\geq 0$

Then, it follows from (\ref{ca}) that

\begin{equation}
E_{3}\geq \lambda \equiv \frac{m_{3}^{2}+A^{2}}{2A}  \label{ela}
\end{equation}%
and it follows from (\ref{f}) that (\ref{ma}) holds. Thus there is an upper
bound on $m_{3}$ but there is no such a bound on $E_{3}$, so extraction of
energy exists and is not restricted! (The reservation is in order that we
work in the test particle approximation, so we neglect backreaction of
particles on the black hole metric). Moreover, we have a lower bound (\ref%
{ela}) instead of the upper one typical of the rotating black hole case \cite%
{p}, \cite{z}, \cite{j}. When $m_{1}\rightarrow 0$, $A\rightarrow 0$ and $%
\lambda \rightarrow \infty $.

IN+  $\varepsilon _{3}=-1$, $C_{1}\geq 0$. 

Then, $F\leq 0$, so we have from (\ref{f}) that%
\begin{equation}
m_{3}\geq A\text{.}
\end{equation}

The conclusion about the lower bound (\ref{ela}) obtained from $C_{1}\geq 0$
now applies as well.

Thus the scenario OUT$^{\_}$ gives no energy extraction and, apart from
this, it forbids creation of superheavy particles. Scenario OUT+ allows
ultra-high energetic particles but with the upper bound on their mass. The
most interesting scenario is IN + since it predicts unbound energies and
superheavy particles with the lower (not upper!) bound on the mass.
Moreover, it is detecticon of superheavy particles that enables to
distinguish between the result of a "standard" Penrose process and its
collisional version in combination with the BSW effect. This is in sharp
contrast with the case of the Kerr metric \cite{p}, \cite{j} and more
general dirty rotating black holes \cite{z}. Thus we have two quite
different situations for the rotating and static charged black hole in what
concerns the combination of the BSW effect and collisional Penrose process.

One reservation is in order. Infalling particle 1 is critical and particle 3
observed at infinity is near-critical. Therefore, if \thinspace $E_{3}\gg
E_{1}$ it means that simultaneously $q_{3}\gg q_{1}$ (if, instead, all
charges have the same order we return to the situation discussed in \cite{jl}
when there is no significant energy extraction). This requires deeply
inelastic collision with participation of composite particles, in which case
it is natural to assume that $q_{3}/q_{1}<137$ as usual. Meanwhile, we would
like to stress that the main result of our work consists in the fact that
enhancement of the observable energy due to the BSW\ effect is possible in
principle, so this can motivate search for more realistic circumstances when
extraction of energy due to the BSW effect can occur. In particular, it is
interesting to consider the possibility of the collisional Penrose process
for the Kerr-Newman metric that combines both opposite limiting cases of the
BSW effect (due to the black hole charge and due to rotation) \cite{kn}.

\begin{acknowledgments}
This work was supported in part by the Cosmomicrophysics section of the
Programme of the Space Research of the National Academy of Sciences of
Ukraine.
\end{acknowledgments}

\end{document}